\documentstyle[nato,numreferences,epsfig]{crckapb}

\newcommand{\preprint}{\vspace{-15.3cm}
 \rightline{\small ITEP-TH-31/99}
 \rightline{\small LU-ITP 1999/022}
 \vspace{1mm}
 \rightline{\small 15 December, 1999}
 \vspace{13.6cm}}
\newcommand{\dubna}{\vspace{-6mm}
 \leftline{\footnotesize{* Presented by the first author 
 at the workshop "Lattice fermions and structure of the}}
 \vspace{-2mm} \leftline{\footnotesize{vacuum", 5-9 Oct. 1999, Dubna,
  Russia.}} \vspace{-3mm}}

\def\beqn{\begin{eqnarray}}
\def\eeqn{\end{eqnarray}}

\newcommand{\eq}[1]{(\ref{#1})}

\begin{opening}
\title{EMBEDDED VORTICES AND THEIR INTERACTIONS \protect\\ 
AT  ELECTROWEAK CROSSOVER$^*$}

\author{M. N. CHERNODUB}
\institute{Institute of Theoretical and Experimental Physics\\
B. Cheremushkinskaja, 25, Moscow, 117259, Russia}
\author{E.--M. ILGENFRITZ}
\institute{Institute of Theoretical Physics,\\
University of Kanazawa, Kanazawa 920-1192, Japan}
\author{A. SCHILLER}
\institute{Institut f\"ur Theoretische Physik \\
Universit\"at Leipzig, D-04109 Leipzig, Germany}

\end{opening}

\runningtitle{EMBEDDED VORTICES AT ELECTROWEAK CROSSOVER}

\begin{document} 
\vspace{-3mm}
\begin{abstract} 
We study properties of $Z$--vortices in
the crossover region of the $3D$ $SU(2)$ Higgs model. 
Correlators of the vortex currents with gauge field 
energy and Higgs field squared 
(``quantum vortex profile'') reveal a structure
that can be compared with a classical vortex. 
We define a core size and a penetration depth from the vortex profile.
$Z$--vortices are found to interact with each
other analogously to Abrikosov vortices in a type--I superconductor. 
\end{abstract}

\dubna

\section{Introduction}

\preprint

Although the standard model does not possess {\it topologically stable}
mo\-no\-po\-le-- and vortex--like defects, 
one can define so-called {\it embedded}
topological defects~\cite{VaBa69,BaVaBu94}: Nambu monopoles~\cite{Na77} and
$Z$--vortex strings~\cite{Na77,Ma83}. In our numerical simulations of the
electroweak 
theory~\cite{ChGuIlSc98-1} we have found that the vortices undergo a
percolation transition which, when there exists a 
discontinuous phase transition at small Higgs masses, 
accompanies the latter.
The percolation transition 
persists at realistic (large) Higgs mass~\cite{ChGuIlSc98-2} 
when the electroweak theory, instead of a transition, 
possesses a smooth crossover around some 
``crossover temperature'' (see Refs.~\cite{Kajantie}). 

We worked in the $3D$ formulation of the $SU(2)$ Higgs model.
This report is restricted
to results obtained in the crossover regime 
(assuming a Higgs boson mass $\approx 100$ GeV).
Details of the lattice model can be found in \cite{generic}.
The defect operators on the lattice have been defined in \cite{ChGuIl98}.
A nonvanishing integer value of the vortex operator $\sigma_P$ on some 
plaquette $P$ signals the presence of a vortex.
The lattice gauge coupling $\beta_G$ is related to the $3D$ continuum gauge 
coupling $g_3^2$ and controls the continuum limit $\beta_G=4/(a g_3^2)$
($g_3^2 \approx g_4^2~T$).
The hopping parameter $\beta_H$ is related to the temperature $T$
(with the higher temperature, symmetric side at 
$\beta_H < \beta_H^{\mathrm{cross}}$). 

\vspace{-3mm}
\section{Vortex profile}
\vspace{-2mm}
Our vortex defect operator $\sigma_P$  
is constructed to localize a line-like object (in $3D$ space--time) 
with non-zero vorticity on the dual lattice. 
Within a given gauge field--Higgs configuration, a profile around 
that vortex ``soul'' would be hidden among quantum fluctuations.
However, an average over all vortices in a quantum ensemble clearly 
reveals a structure that can be compared with a classical 
vortex~\cite{Na77,BaVaBu94}. 
We have studied correlators of $\sigma_P$ with various operators 
constructed on the lattice (``quantum vortex profiles'').

\begin{figure}
  \begin{minipage}{6.0cm}
    \hfill
     \epsfig{file=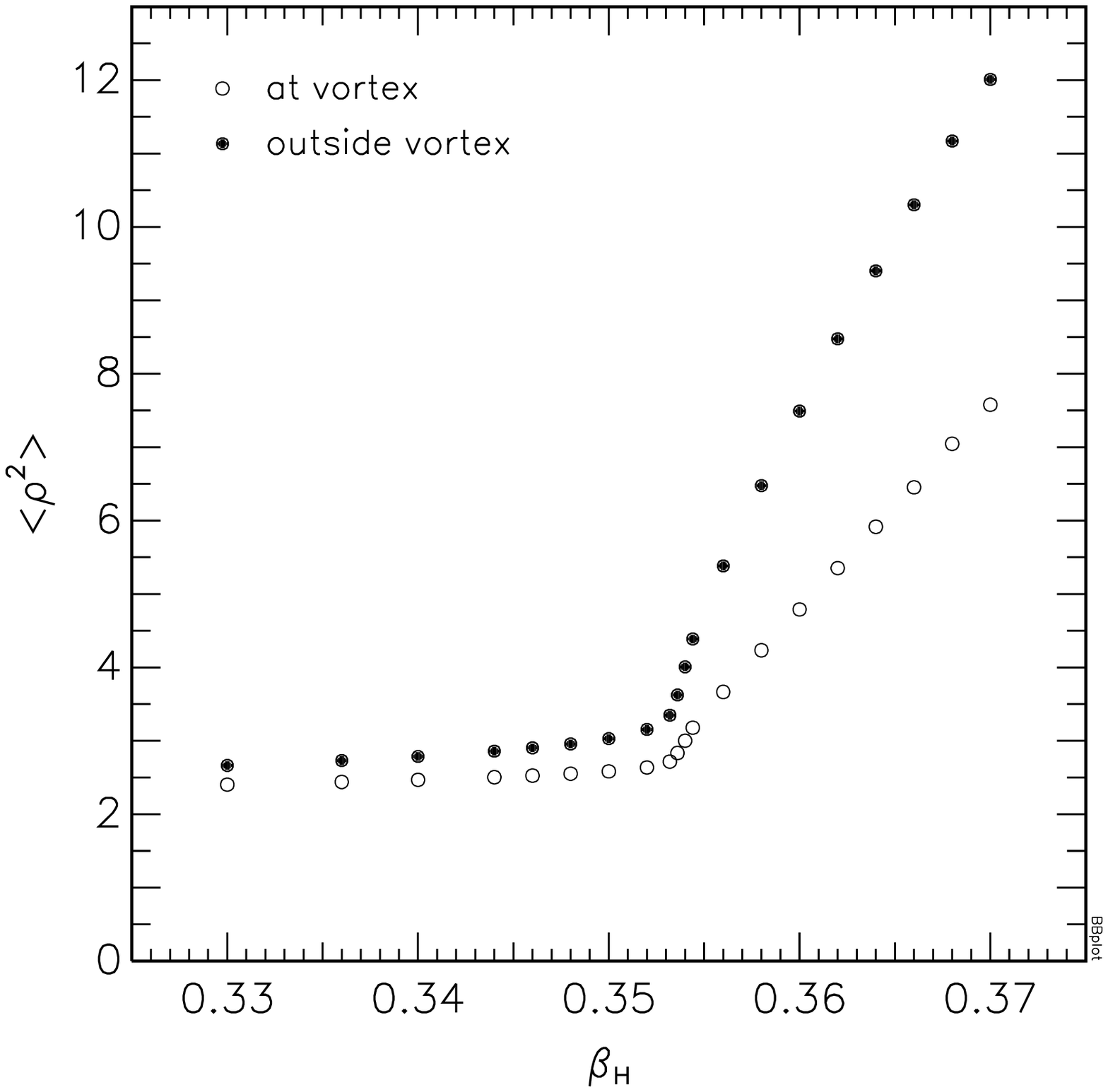,
      height=5.0cm,width=5.0cm}
  \end{minipage}
  \hfill
  \begin{minipage}{6.0cm}
     \hfill
     \epsfig{file=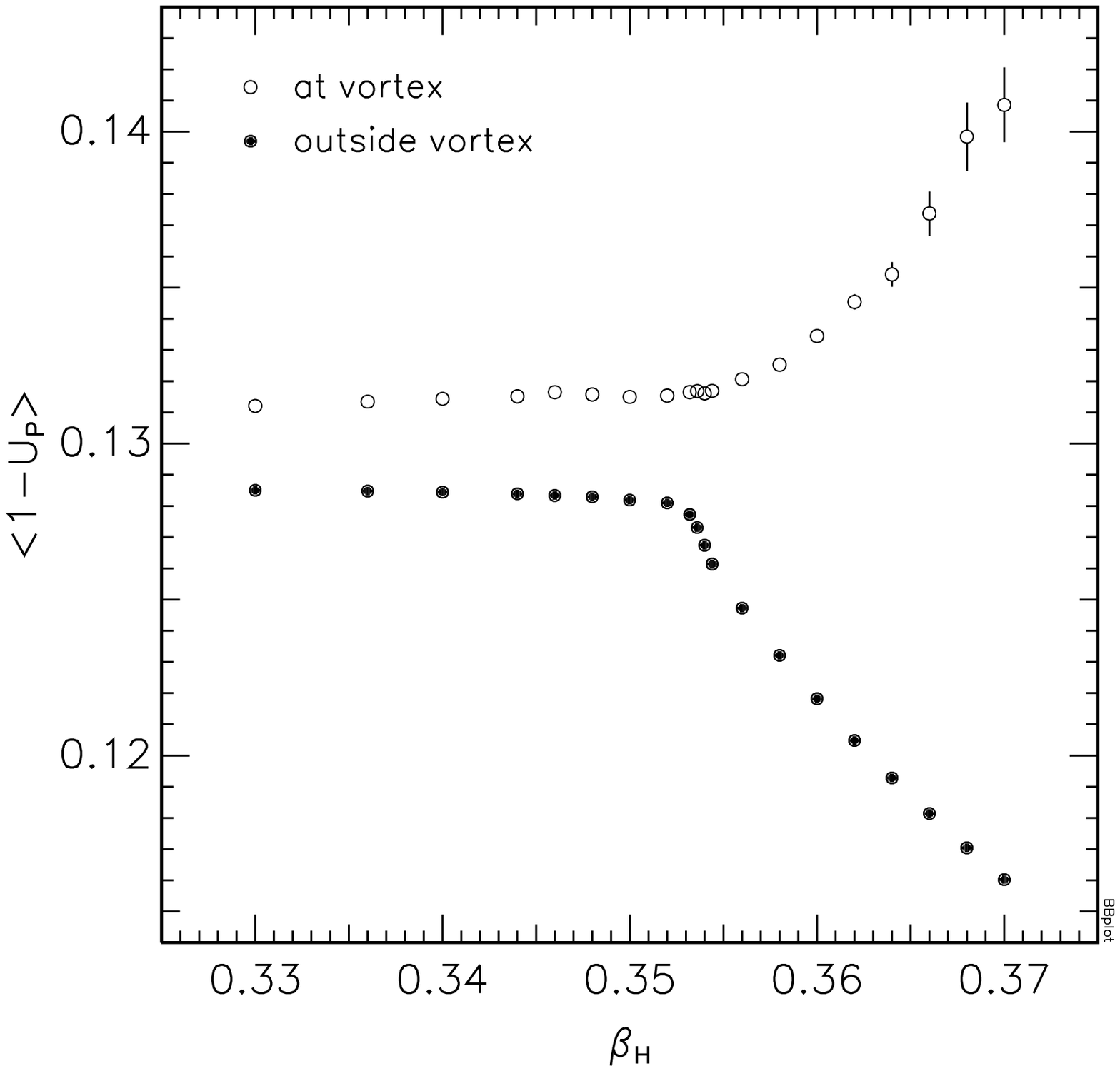,
      height=5.0cm,width=5.0cm}
    \end{minipage}
\vspace{-2mm}
 \caption{
Higgs modulus squared and gauge field energy inside and 
outside of a vortex  {\it vs.} $\beta_H$, $\beta_G=8$.}
\label{rhosg}
\vspace{-5mm}
\end{figure}
Classically, in the center of a vortex the Higgs field modulus turns to 
zero and the energy density becomes maximal~\cite{Na77,BaVaBu94}.  
What can be expected in a thermal ensemble is, that along the vortex soul 
the (squared) modulus of the Higgs field and the gauge field energy density,
$E^g_P = 1-\frac12 {\mathrm{Tr}} U_P$, substantially differ   
from the bulk averages characterizing the corresponding homogeneous 
phase.\footnote{Just on the ``broken'' side of the crossover, for instance, 
one would expect to find a core of ``symmetric'' matter inside the vortex.}
Indeed, in our lattice study they were found lower (or higher, respectively),
with the difference growing entering deeper into the
``broken phase'' side of the crossover~\cite{ChGuIlSc98-2}
(see Figure~\ref{rhosg}).

To proceed we have studied, among others, the vortex--gluon energy 
correlator for plaquettes $P_0$ and $P_R$ located in the same plane 
(perpendicular to a segment of the vortex path)  
\beqn
C_E(R)    =  \langle \sigma^2_{P_0} \, E^g_{P_R} \rangle \,,
\label{CorrDefs}
\eeqn
as function of the distance $R$ between the 
plaquettes.\footnote{A similar method has been used to study the 
physical properties of Abelian monopoles in
$SU(2)$ gluodynamics, Ref.~\cite{MIPHarald}.} 
\begin{figure}
 \begin{minipage}{12.5cm}
 \begin{center}
  \epsfig{file=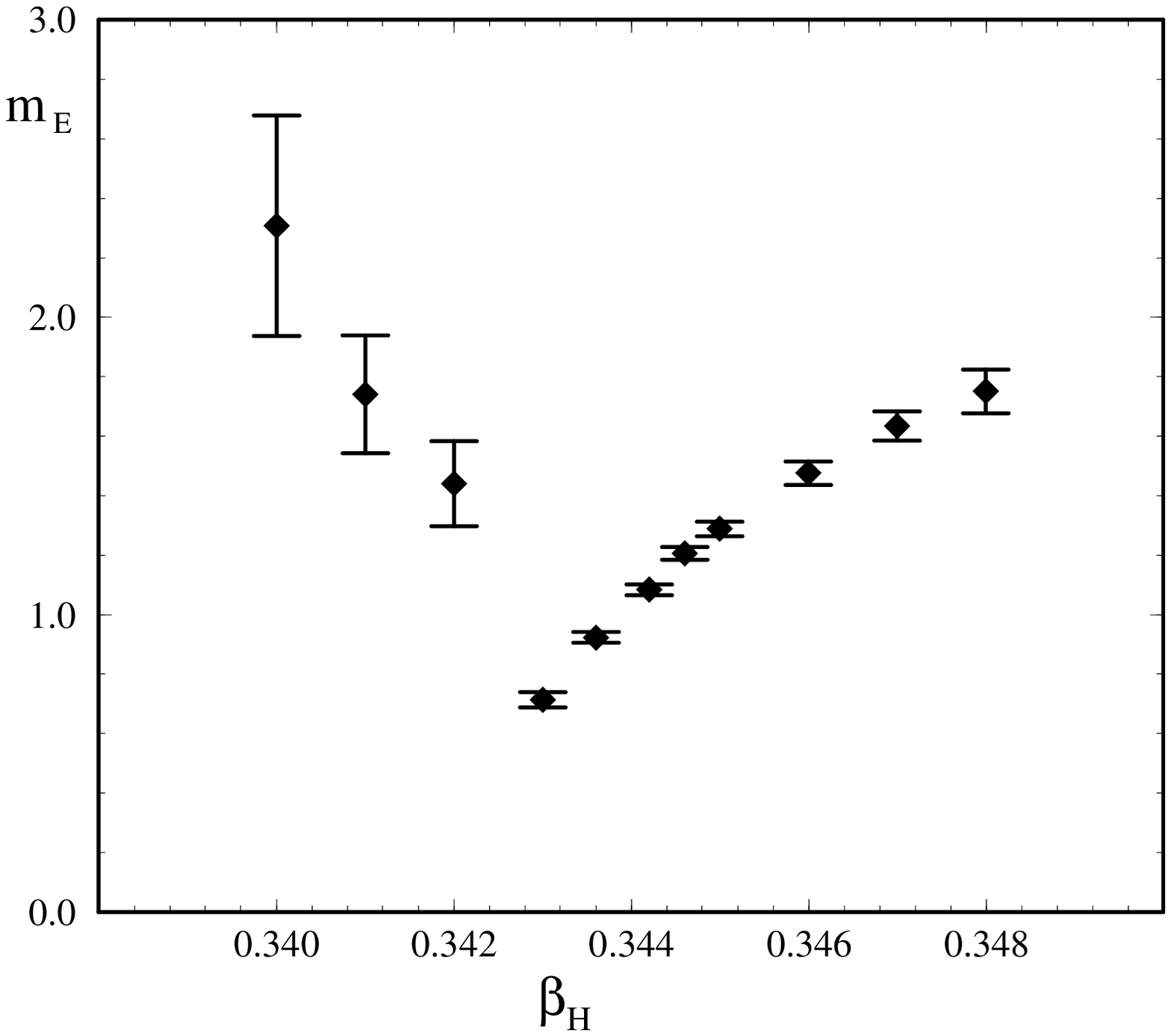,width=5.8cm,height=5.0cm} \hspace{5mm}
  \epsfig{file=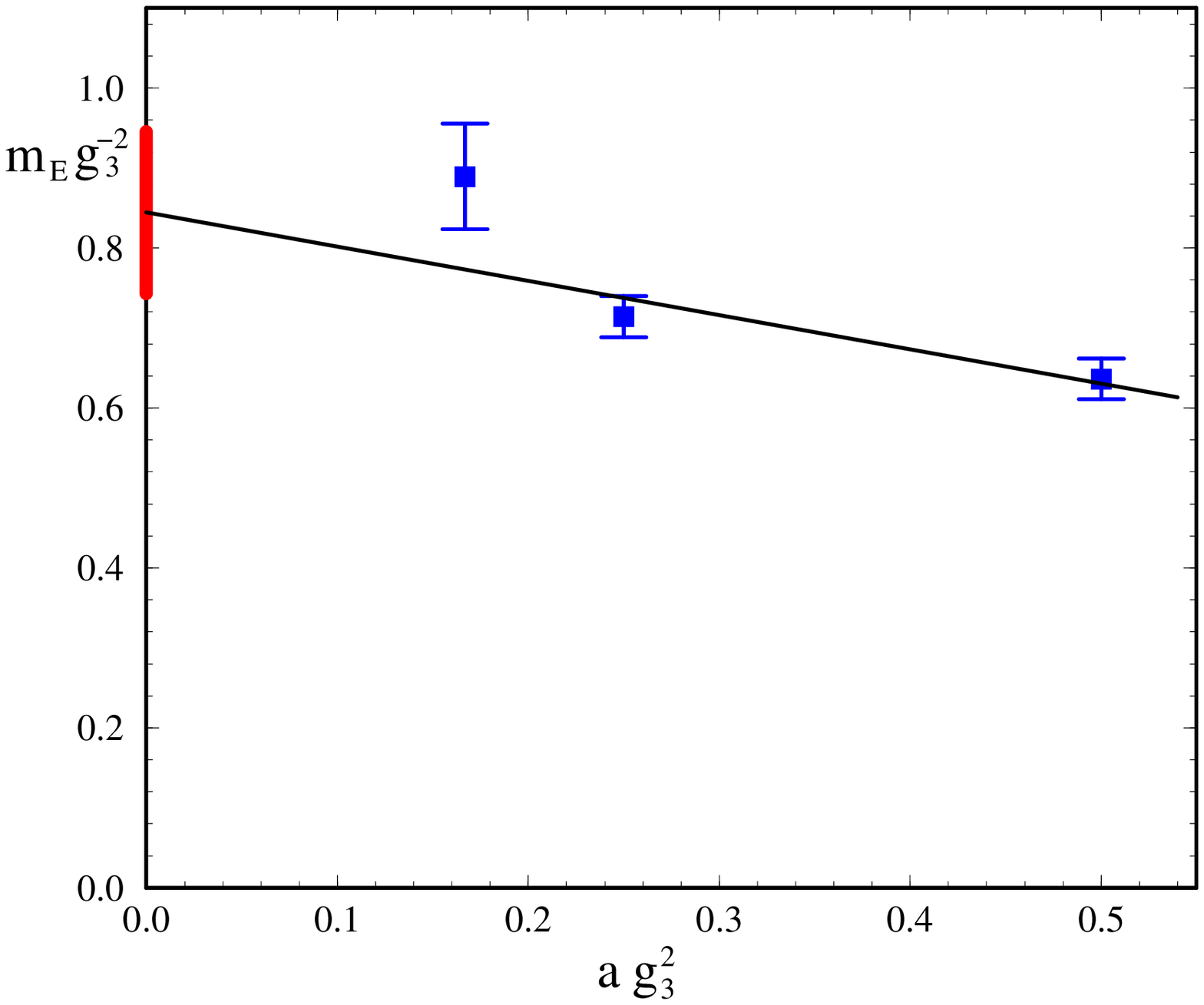,width=5.8cm,height=5.0cm}\\
  \vspace{-2mm}
(a) \hspace{5.5cm} (b)\\
 \end{center}
 \end{minipage}
 \caption{(a) The effective mass $m_E$ {\it vs.} hopping parameter $\beta_H$
at $\beta_G=16$ on the lattice $32^3$; (b)
Extrapolation of the mass $m_E$ fitted at crossover to the limit $a \to 0$.}
\label{vortexEmass}
\end{figure}
To parametrize the vortex shape we fit the correlator data \eq{CorrDefs} by 
an ansatz $C^{\mathrm{fit}}_E(R) = C_E + B_E~G(R;m_E)$ with 
constants $C_E$ and $B_E$ and an inverse penetration depth 
(effective mass $m_E$). The function 
$G(R;m)$ is the $3D$ scalar lattice propagator with mass $2 \sinh (m/2)$
which, instead of a pure exponential, has been proposed to fit 
{\it point--point} correlators in Ref.~\cite{Engels:1995ek}. 

If the quantum vortex profile 
should interpolate between the interior of the vortex 
and the asymptotic approach to the vacuum, we can only
expect to describe the profile by such an ansatz 
for distances $R>R_{\mathrm{min}}$.
The distance $R_{\mathrm{min}}$ (core size) should be fixed in physical 
units. Therefore we choose (in lattice units) 
$R_{\mathrm{min}}(\beta_G)$=$\beta_G \slash 8$ for $\beta_G$=8,16,24 
which corresponds to
$R^{\mathrm{core}}$=$a R_{\mathrm{min}}$=${(2 g^2_3)}^{-1}$. 
How successful this is to define the vortex core can be assessed studying 
$\chi^2 \slash d.o.f.$ {\it vs.} $R_{\mathrm{min}}$ 
(to be reported elsewhere).

An example of the behaviour of the effective mass $m_E$ 
is shown in Figure~\ref{vortexEmass}(a). The mass reaches its
minimum at the crossover point $\beta_H^{\mathrm{cross}}$.  
Deeper on the symmetric side 
the quantum vortex profiles are squeezed compared to the classical ones 
due to Debye screening leading to a smaller coherence length.  
Approaching the crossover from this side the density of the vortices 
decreases thereby diminishing this effect.
The extrapolation of the mass $m_E$ (as defined at the crossover temperature)
towards the continuum limit is shown in Figure~\ref{vortexEmass}(b). 

\section{Inter--vortex interactions and the type of
the vortex medium}
\vspace{-2mm}
In the case of a superconductor, the inter--vortex
interactions define the type of superconductivity. 
If two parallel static vortices with the same sense of vorticity attract
(repel) each other, the substance is said to be a type--I (type--II)
superconductor. To investigate the vortex--vortex interactions 
we have measured two--point functions of the vortex currents:
\beqn
\langle|\sigma_{P_0}|\,|\sigma_{P_R}|\rangle=2 
   (g_{++}+g_{+-}) \,, \ \
\langle \sigma_{P_0}\,\sigma_{P_R}\rangle =2 
   (g_{++}-g_{+-}) \,, 
\eeqn
where $g_{+\pm}(R)$ stands for contributions to the correlation 
functions from parallel/anti--parallel  vortices 
piercing a plane in plaquettes $P_0$ and $P_R$. 
Properly normalized, the correlators $g_{+\pm}(R)$ 
can be interpreted as the average density of vortices (anti-vortices),
relative to the bulk density, at distance $R$ from a given vortex. 

Hence the long range tail of the function $g_{++}$ is crucial 
for the type of the vortex medium: 
in the case of attraction (repulsion) between same sign vortices 
$g_{++}$ exponentially approaches unity from above (be\-low)
while $g_{+-}$ is always attractive,  
independently on the type of superconductivity. 

We have seen in our calculations \cite{lat99}
that the tail of $g_{++}$ belongs to the attraction case (with
minimal slope at the crossover). Therefore, electroweak matter in 
the crossover regime belongs to the type--I vortex vacuum class.

\vspace{-4mm}
\section*{Acknowledgments}
\vspace{-3mm}
The authors are grateful to P.~van~Baal, H.~Markum, V.~Mitrjushkin, 
S. Olejnik and M.~I.~Polikarpov for useful discussions. 
M.~Ch. feels much obliged for the kind hospitality
extended to him at the Max-Planck-Institute for Physics in Munich. 
M.~Ch. was supported by the
grants INTAS-96-370, RFBR-99-01-01230 and ICFPM fellowship (INTAS-96-0457).


\vspace{-3mm}
\begin{thebibliography}{99}
\vspace{-2mm}
\bibitem{VaBa69}
T.~Vachaspati and M.~Barriola, {\it Phys. Rev. Lett.}
{\bf 69} (1992) 1867.

\bibitem{BaVaBu94} M.~Barriola, T.~Vachaspati, M.~Bucher,
{\it Phys.~Rev.} {\bf D50} (1994) 2819.

\bibitem{Na77}
Y.~Nambu, {\it Nucl.~Phys.} {\bf B130} (1977) 505.

\bibitem{Ma83}
N.~S.~Manton, {\it Phys.~Rev.} {\bf D28} (1983) 2019.

\bibitem{ChGuIlSc98-1}
M.~N.~Chernodub  {\it et al.}, {\it Phys.~Lett.} {\bf B434} (1998) 83.

\bibitem{ChGuIlSc98-2}
M.~N.~Chernodub  {\it et al.}, {\it Phys.~Lett.} {\bf B443} (1998) 244.

\bibitem{Kajantie}
K.~Kajantie {\it et al}, {\it Phys. Rev. Lett.} {\bf 77}
(1996) 2887; M.~G\"urtler, E.--M.~Ilgenfritz, A.~Schiller,
{\it Phys.~Rev.} {\bf D56} (1997) 3888.

\bibitem{generic} K.~Kajantie {\it et al.}, {\it Nucl. Phys.} {\bf
B458} (1996) 90; M.~G\"urtler {\it et al.}, {\it ibid.} {\bf B483} (1997) 383.

\bibitem{ChGuIl98}
M.~N.~Chernodub, F.~V.~Gubarev, E.--M.~Ilgenfritz,
{\it Phys.~Lett.} {\bf B424} (1998) 106.

\bibitem{MIPHarald} 
S. Thurner {\it et al.}, {\it Phys.~Rev.} {\bf D} 54 (1996) 3457;
M.~Feurstein, H.~Markum and S. Thurner, {\it Phys.Lett.} {\bf B396} (1997) 203;
B.~L.~G.~Bakker, M.~N.~Chernodub, M.~I.~Polikarpov,
{\it Phys.~Rev.~Lett.} {\bf 80} (1998) 30.

\bibitem{Engels:1995ek} J.~Engels, V.~K.~Mitryushkin, T.~Neuhaus,
{\it Nucl.~Phys.} {\bf B440} (1995) 555.

\bibitem{IlScSt99}
E.--M.~Ilgenfritz, A.~Schiller and C.~Strecha, 
{\it Eur.~Phys.~J.} {\bf C8} (1999) 135.

\bibitem{lat99}
M.~N.~Chernodub, E.--M.~Ilgenfritz and A.~Schiller, heplat/9909001.

\end{thebibliography}
\end{document}